\documentclass[12pt]{article}
\usepackage{graphics}
\begin{document}
\title{Magnetic moments of octet baryons, angular momenta of quarks
 and sea antiquark polarizations}
\author{ Jan
Bartelski\\ Institute of Theoretical Physics,\\ Faculty of Physics,
Warsaw University,\\ Ho$\dot{z}$a 69, 00-681 Warsaw, Poland. \\ \\
\and Stanis\l aw Tatur
\\ Nicolaus Copernicus Astronomical Center,\\ Polish Academy of
Sciences,\\ Bartycka 18, 00-716 Warsaw, Poland. \\ }
\date{}
\maketitle
\begin{abstract}
\noindent One can determine antiquark polarizations in a proton
using the information from deep inelastic scattering, $\beta$ decays
of baryons, orbital angular momenta of quarks, as well as their
integrated magnetic distributions. The last quantities were
determined previously by us performing a fit to magnetic moments of
a baryon octet. However, because of the $SU(3)$ symmetry our results
depend on two parameters. The quantity $\Gamma_{V}$, measured
recently in a COMPASS experiment, gives the relation between these
parameters. We can fix the last unknown parameter using the ratio of
up and down quark magnetic moments which one can get from the fit to
radiative vector meson decays. We calculate antiquark polarizations
with the orbital momenta of valence quarks that follow from lattice
calculations. The value of the difference of up and down antiquark
polarizations obtained in our calculations is consistent with the
result obtained in a HERMES experiment.
\end{abstract}

\newpage
\section{Introduction and Framework}
In Ref. \cite{jb} we have proposed a model for magnetic moments of
SU(3) octet baryons. We get an excellent fit using few parameters.
In this model magnetic moments of baryons are sums of products of
magnetic moments of quarks and corresponding integrated quark
densities. The corrections, which take into account exchange
phenomena, were also included. To determine the size of such
corrections we use sum rules for magnetic moments of octet baryons
(as in \cite{jb} and \cite{jbr}). After subtraction of the exchange
contributions we are left with a $SU(3)$ symmetric part of these
moments, which can be expressed as independent contributions from
quarks and antiquarks in considered baryon. The magnetic moment of a
quark as well as its integrated magnetic density are $Q^{2}$
dependent, whereas its product is not. Other corrections coming from
exchange effects of pions and gluons are incorporated in
redefinition magnetic moments of quarks and their integrated
densities; hence magnetic moments of quarks are not equal to their
Dirac values.

So, after subtracting  the pion correction to nucleon magnetic
moments and taking into account $\Sigma^{0}-\Lambda $ mixing, we are
left with independent one particle contributions to baryon magnetic
moments (sum rules for magnetic moments are satisfied) and we use
for them high energy parametrization (integrated parton densities)
to describe such contributions. We believe that most of all other
pion exchange and gluon exchange corrections are taken into account
in the high energy parametrization (see, e.g., \cite{jbr1}).  For
such parametrization, in the case of axial densities, one does not
include explicitly pion and gluon corrections and we do the same for
integrated magnetic densities. In contrary such corrections are
present in models of bound quarks, e.g., in \cite{bhy}, \cite{thom3}
and \cite{myh}. One gluon correction with gluon exchanged between
different quarks can also correspond to higher twist diagrams in
deep inelastic scattering.

In integrated magnetic quark densities, besides of spin
contributions, we have also orbital angular momentum contributions
(see also \cite{ju}).
 Here we shall consider two models: the one in which we
neglect orbital angular momentum contribution and the second with
such contribution included. In the first case we have for such
integrated densities

 \begin{equation}
\delta q \equiv \Delta q_{val}+\Delta q_{sea}-\Delta\bar{q},
\end{equation}

In the case with angular momentum the formulas are

\begin{equation}
\delta_{L} q \equiv \Delta q_{val}+\Delta
q_{sea}-\Delta\bar{q}+L_{q},
\end{equation}

where

\begin{equation}
L_{q}=<\hat{L}_{z}^q>-<\hat{L}_{z}^{\bar
q}>=<\hat{L}_{z}^{q_{val}}>+<\hat{L}_{z}^{q_{sea}}>-<\hat{L}_{z}^{{\bar
q}_{sea}}>.
\end{equation}

Taking into account exchange contributions (as was explained in
detail in [1]), i.e., isovector contribution connected with charged
pion exchange between different quarks (see also Franklin
\cite{fr1,fr2}) and $\Sigma^{0}-\Lambda $ mixing, the $SU(3)$
symmetric part of baryon octet  magnetic moments can be parametrized
in terms of four quantities: $c_{0}$, $c_{3}$, $c_{8}$, $r$. From
the fit we get for these parameters \cite{jb}
\begin{eqnarray}
c_{0} &= & 0.054 \pm 0.001\, n.m. \, , \nonumber \\
c_{3} &=& 1.046 \pm 0.005\, n.m. \, , \nonumber \\
c_{8} & =& 0.193 \pm 0.000\, n.m. \, , \\
r & =& 1.395 \pm 0.010\,   \nonumber.
\end{eqnarray}
Hence, six quantities: three quark magnetic moments and three quark
densities cannot be determined using only four parameters given in
Eq. (4). So as in [1], we introduce two additional parameters,
$\epsilon$ and $g$, and our quantities become the functions of them.
The parameters $\epsilon$ and $g$ are defined in Eqs. (5) and (6):

\begin{equation}
\epsilon =-1-2\frac{\mu_{d}}{\mu_{u}},
\end{equation}

\begin{equation}
g=\delta_{L} u -\delta_{L} d .
\end{equation}

Now we can express magnetic quark densities as
\begin{eqnarray}
\delta_{L} u &=& \frac{g}{6r}[f(\epsilon)+1+3r] \, , \nonumber \\
\delta_{L} d &=& \frac{g}{6r}[f(\epsilon)+1-3r] \, , \\
\delta_{L} s &=& \frac{g}{6r}[f(\epsilon)-2] \, ,\nonumber
\end{eqnarray}
where

\begin{equation}
 f(\epsilon)=\frac{(3+\epsilon) r c_{0}}{c_{3}-3rc_{8}-\epsilon (c_{3}+rc_{8})} \,
 .
\end{equation}

 One can also express magnetic moments of u, d and s quarks in
terms of our parameters $\epsilon$ and $g$:
\begin{eqnarray}
\mu_{u} &=&  \frac{8 c_3}{g(3+\epsilon)}\, , \nonumber \\
\mu_{d} &=& -\frac{4 (1+\epsilon) c_3 }{g(3+\epsilon)} \, , \\
\mu_{s} &=& -\frac{2[9 r c_8- c_3
+\epsilon(c_3+3rc_8)]}{g(3+\epsilon)} \, . \nonumber
 \end{eqnarray}

 The parameter $g$ sets a scale at which we have calculated our
 quantities. From Eqs. (7) and (9) we have that $\mu_{q}\delta_{L} q$
 and quark magnetic moment ratios, e.g., $\mu_{u}/\mu_{d}$, do not
 depend on $g$.

The new quantity $\Gamma_{V}$, which in our notation is

\begin{equation}
\Gamma_{V} \equiv  \delta u+\delta d=\delta_{L} u+ \delta_{L}
d-L_{u}-L_{d} ,
\end{equation}

is measured in the COMPASS experiment \cite{comp} and one gets

\begin{equation}
\Gamma_{V}=0.41 \pm 0.07(stat.) \pm 0.06(syst) \, .
\end{equation}

In the case when $ \Delta q_{sea} \neq \Delta \bar{q}$ this quantity
is not a valence one: $ \Delta u_{val}+\Delta d_{val}$.

Using Eqs. (7) and (10) we can express our parameter $g$ as:

\begin{equation}
g=\frac{3r( \Gamma_{V}+ L_{u} + L_{d})}{f(\epsilon)+1}.
\end{equation}

Hence, the  COMPASS measurement gives the relation between
introduced parameters $\epsilon$ and $g$. So we will have only one
unknown parameter however, the orbital angular momenta of quarks are
present in the formulas.

We know that integrated axial densities, used in deep inelastic
scattering analysis, differ from $\delta{q}$ by a sign in an
antiquark term:

\begin{equation}
\Delta q \equiv \Delta q_{val}+\Delta q_{sea}+\Delta\bar{q} \, .
\end{equation}

From Eqs. (1,2) and (13) we can express $\Delta\bar{q}$ as

\begin{equation}
\Delta\bar{q}= \frac{1}{2}(\Delta q-\delta q) = \frac{1}{2}(\Delta
q-\delta_{L} q + L_{q}).
\end{equation}

Let us express the function $f(\epsilon)$ in the form

\begin{equation}
f(\epsilon)=\frac{3r( \Gamma_{v}+ L_{u} + L_{d})}{a_{3}-2\eta+
L_{u}-L_{d}}-1,
\end{equation}

where $\eta$ is defined by
\begin{equation}
\eta \equiv \Delta\bar{u} - \Delta\bar{d}.
\end{equation}

Equation (15) gives us a relation between our two basic parameters
$\epsilon$ and $\eta$ (which replaces parameter $g$). The quark
integrated axial densities $\Delta u $, $\Delta d $, $\Delta s $ can
be determined from
\begin{eqnarray}
 \Delta u &=& \frac{1}{3}a_{0}+ \frac{1}{6}a_{8}+ \frac{1}{2}a_{3}\, , \nonumber \\
 \Delta d &=& \frac{1}{3}a_{0}+ \frac{1}{6}a_{8}- \frac{1}{2}a_{3} \, , \\
 \Delta s &=& \frac{1}{3}a_{0}- \frac{1}{3}a_{8} \, ,\nonumber
\end{eqnarray}
where the values of $a_{3}$, $a_{8}$ and $a_{0}$ are obtained from
neutron and hyperon $\beta$ decays \cite{pdg} and deep inelastic
scattering spin experiments \cite{comp1}. We have
\begin{eqnarray}
a_{0} &=& 0.33 \pm 0.06 \, , \nonumber \\
a_{8} &=& 0.585 \pm 0.025 \, , \\
a_{3} &=& 1.2694 \pm 0.0028 \, .\nonumber
\end{eqnarray}

We can get additional information (although not very precise) using
the value of $\eta$ from the HERMES experiment \cite{her}. We will
take $\eta = 0.05 \pm 0.06$ which was however measured not in the
whole range of $x$: ($0.023 \leq x \leq 0.6 $).

Hence we can express $\Delta \bar{u}$, $\Delta \bar{d}$ and $\Delta
\bar{s}$ as a function of parameter $\eta$ using Eqs. (7), (12),
(15), and (17), getting

\begin{eqnarray}
 \Delta \bar{u} &=& \frac{1}{6}a_{0}+ \frac{1}{12}a_{8}- \frac{1}{4}\Gamma_{V}+\frac{1}{2} \eta\, , \nonumber \\
 \Delta \bar{d} &=& \frac{1}{6}a_{0}+ \frac{1}{12}a_{8}- \frac{1}{4}\Gamma_{V}-\frac{1}{2} \eta \, , \\
 \Delta \bar{s} &=& \frac{1}{6}a_{0}- \frac{1}{6}a_{8}- \frac{1}{4}\Gamma_{V}+\frac{a_{3}-2\eta}{4r}+
 \frac{L_{u}-L_{d}}{4r} -\frac{L_{u}+L_{d}-2L_{s}}{4} \, .\nonumber
\end{eqnarray}

One sees that $\Delta \bar{u}$, and $\Delta \bar{d}$ do not depend
directly on orbital angular momenta. Knowing the precise value of
$\eta$ in the whole range ($0 \leq x \leq 1 $) one is able to
determine $\Delta \bar{u}$ and $\Delta \bar{d}$ in our model.
\section{Numerical results and discussion}
\subsection{Angular momenta of quarks neglected}

\begin{figure}
\noindent \scalebox{0.8}{ \includegraphics{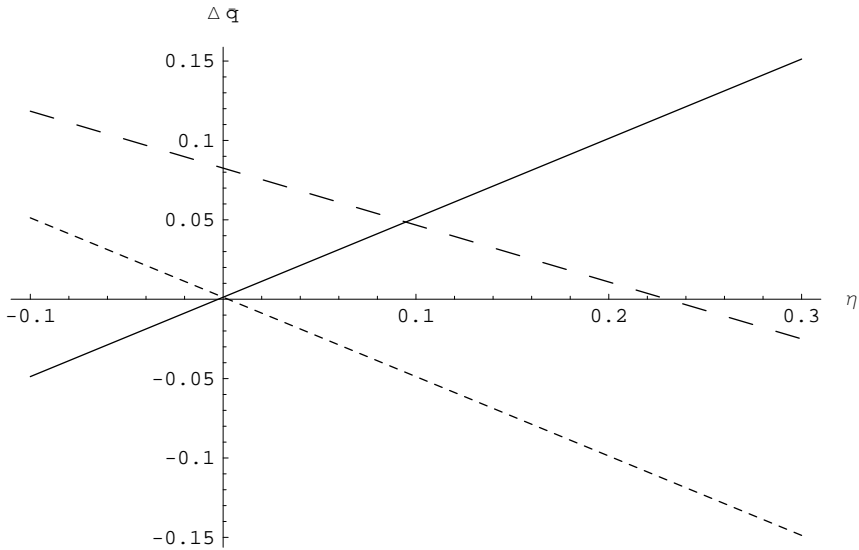}}

\caption{\label{fig1}\textit{The antiquark polarizations for
$\bar{u}$ (solid line), $\bar{d}$ (short-dashed line), and $\bar{s}$
(long-dashed line) versus $\eta$ in the model where angular momenta
of quarks are neglected.}}
\end{figure}

Let us start with an assumption that all angular momenta of quarks
are negligible (i.e., we put them equal to zero). In Fig.~\ref{fig1}
we present antiquark polarizations $\Delta \bar{q}$ for u, d and s
quarks as a functions of parameter $\eta$.

\begin{figure}
\noindent \scalebox{0.8}{ \includegraphics{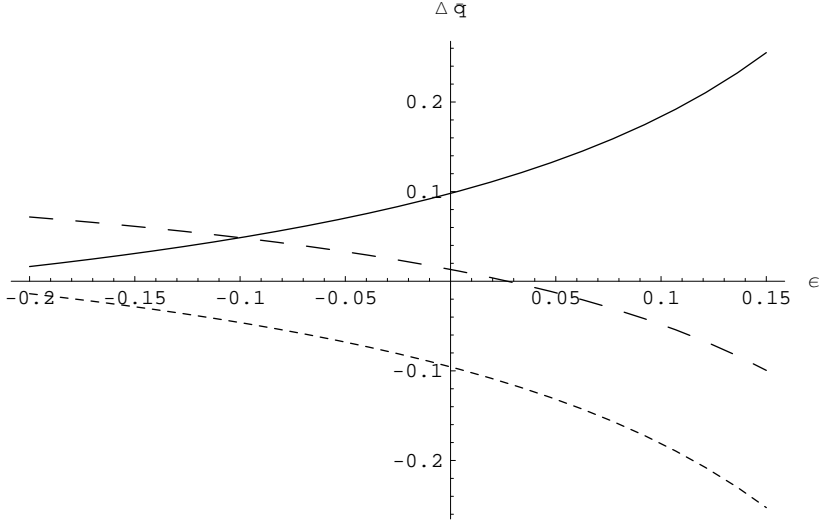}}

\caption{\label{fig2}\textit{The antiquark polarizations for
$\bar{u}$ (solid line), $\bar{d}$ (short-dashed line), and $\bar{s}$
(long-dashed line) versus $\epsilon$ in the model where angular
momenta of quarks are neglected.}}
\end{figure}

The fact that $\Delta \bar{u}+\Delta \bar{d} \approx 0$ is connected
with the value of $\Gamma_{V}$ measured by the COMPASS experiment.
We can also see how the values of $\Delta \bar{u}$, $\Delta
\bar{d}$, and $\Delta \bar{s}$ change when we change $\eta$ between
$-0.01$ and $0.11$, i.e., within 1 standard deviation off central
value. When we take the number from the HERMES experiment ($\eta =
0.05$), we can determine polarizations of all sea antiquarks:
\begin{eqnarray}
\Delta \bar{u} &=& 0.03 \pm 0.04  \, , \nonumber \\
\Delta \bar{d} &=& -0.02 \pm 0.04  \, , \\
\Delta \bar{s} &=&  0.06 \pm 0.03 \, .\nonumber
\end{eqnarray}
The values are a little bit different from the antiquark values
quoted by HERMES \cite{her}. Using Eqs. (4), (7) and (15) we can
calculate the corresponding value of parameter $\epsilon$. We get
$\epsilon = - 0.17 $ for $\eta = 0.05$. In general we can use Eq.
(15) to eliminate $\eta$ and express $\Delta \bar{q}$ as a function
of $\epsilon$ in the case when we neglect dependence on orbital
angular momenta. In Fig.~\ref{fig2} we show such dependence, i.e.,
$\Delta \bar{q}(\epsilon)$.

One can try to determine the value of parameter $\epsilon$ using the
experimental data for radiative vector meson decays. The model which
is used to determine $\frac{\mu_{u}}{\mu_{d}}$ is not as
sophisticated as is the one for baryon magnetic moments; we have
used similar formulas as in \cite{vec}. One does not include the
contribution from orbital momenta of quarks in such a model;
however, in \cite{gloz} it was shown that such contributions may be
small. Performing the fit one gets $\frac{\mu_{u}}{\mu_{d}} = - 1.87
\pm 0.07$ which gives, with the help of Eq. (5),   $\epsilon = 0.06
\pm 0.04$. If we use this value of parameter $\epsilon$ we will get
for $\Delta \bar{q}$
\begin{eqnarray}
\Delta \bar{u} &=& 0.14 \pm 0.07  \, , \nonumber \\
\Delta \bar{d} &=& -0.14 \pm 0.04  \, , \\
\Delta \bar{s} &=&  -0.02 \pm 0.03 \, .\nonumber
\end{eqnarray}
In this case the corresponding value of parameter $\eta$ is $0.28
\pm 0.08$. It looks as if $\Delta \bar{q}$ calculated from the
HERMES value of $\eta$ and $\Delta \bar{q}$ calculated using
$\epsilon$ gotten from vector meson decays are not consistent.
\subsection{Angular momenta of quarks taken into account}
\begin{figure}
\noindent \scalebox{0.8} { \includegraphics{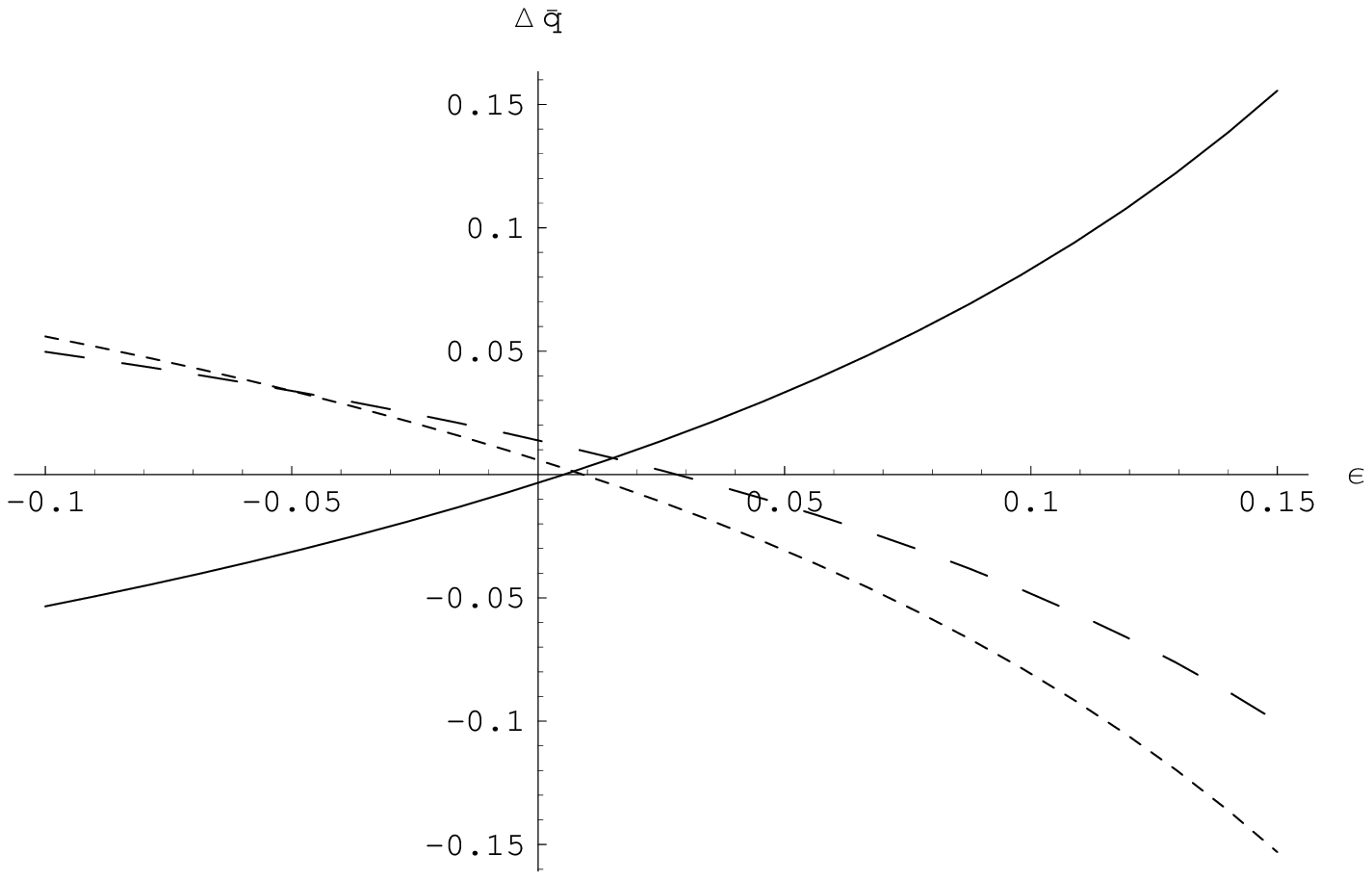}}

\caption{\label{fig3}\textit{The antiquark polarizations for
$\bar{u}$ (solid line), $\bar{d}$ (short-dashed line), and $\bar{s}$
(long-dashed line) versus $\epsilon$ in the model where angular
momenta of quarks are taken into account.}}

\end{figure}

Now we shall consider the model with nonzero orbital angular momenta
of quarks. We will use the values of such momenta calculated
numerically on the lattice. From \cite{hag} we have
\begin{eqnarray}
L_{u} &=& L_{u}^{val}(lattice)=-0.195 \pm 0.044 \, , \\ L_{d} &=&
L_{d}^{val}(lattice)=0.200 \pm 0.044 \, .\nonumber
\end{eqnarray}
These values are determined at $Q^{2}=4\; GeV^2$. Let us make some
comments. From Eq. (3) angular momentum of quarks consists of
angular momentum of valence quarks, sea quarks, and antiquarks. From
\cite{hag} we have only information on valence quark contribution.
We neglect the rest because of lack of knowledge; it actually means
that we assume that orbital angular momenta of sea quarks and
antiquarks are equal [see Eq. (3)]. The existence of a small
correction to this hypothesis cannot be excluded. The orbital
angular momentum of quarks is scale dependent \cite{Ji}. There are
two possibilities: First one can take into account evolution
equations for angular momenta and start with initial conditions at
low energies taking as is suggested by A. W. Thomas values that
follow from the cloudy bag model \cite{thom1}, \cite{thom2}, which
take into account the relativistic motion of quarks, chiral pion
cloud, and one gluon exchange corrections. Second, one can take
initial conditions at high energy as was done in \cite{wak}. In our
case we use high energy parameters so it is natural to use high
energy initial conditions as in \cite{wak}. From our procedure it
seems that we cannot go in $Q^{2}$ scale below $1\; GeV^2$. From
\cite{wak} it follows that the angular momenta of quarks for
$Q^{2}>1\; GeV^2$ are only weakly scale dependent. The angular
momenta of valence quarks and $\Gamma_{V}$ determine the scale used
in our equations. Using Eq. (19) and eliminating $\eta$ [using Eq.
(15)] we can get formulas for $\Delta \bar{q}$ (for u, d and s
quarks) with orbital angular momenta taken into account:
\begin{eqnarray}
 \Delta \bar{u} &=& \frac{1}{6}a_{0}+ \frac{1}{12}a_{8}+\frac{1}{4}a_{3}- \frac{1}{4}\Gamma_{V}-
 \frac{3r}{4} \frac{\Gamma_{V}+L_{u}+L_{d}}{f(\epsilon)+1}+\frac{1}{4}(L_{u}-L_{d}) \, , \nonumber \\
 \Delta \bar{d} &=& \frac{1}{6}a_{0}+ \frac{1}{12}a_{8}-\frac{1}{4}a_{3}-
 \frac{1}{4}\Gamma_{V}+
 \frac{3r}{4} \frac{\Gamma_{V}+L_{u}+L_{d}}{f(\epsilon)+1}-\frac{1}{4}(L_{u}-L_{d}) \, , \\
 \Delta \bar{s} &=& \frac{1}{6}a_{0}- \frac{1}{6}a_{8}-
 \frac{1}{4}\Gamma_{V}+
 \frac{3r}{4} \frac{\Gamma_{V}+L_{u}+L_{d}}{f(\epsilon)+1}-\frac{1}{4}(L_{u}+L_{d}-2L_{s})  \, .\nonumber
\end{eqnarray}
The dependence of $\Delta \bar{q}$ on $\epsilon$, calculated from
Eq. (23), is shown in Fig.~\ref{fig3}.

When we use the result for $\epsilon$ from the fit to radiative
decays of vector mesons we get for $\Delta \bar{u}$, $\Delta
\bar{d}$ and $\Delta \bar{s}$

\begin{eqnarray}
\Delta \bar{u} &=& 0.04 \pm 0.08  \, , \nonumber \\
\Delta \bar{d} &=& -0.04 \pm 0.05  \, , \\
\Delta \bar{s} &=&  -0.02 \pm 0.03 \, .\nonumber
\end{eqnarray}

The errors are quite big so the determination is not very
conclusive. For $\eta = \Delta \bar{u} - \Delta \bar{d}$ we get the
value $0.08 \pm 0.09$ which has to be compared with $0.05 \pm 0.06$.
The agreement is reasonable despite the fact that all errors are
relatively big. Let us stress that we have used the value of
$\epsilon$ calculated from the fit to experimental data on radiative
vector meson decays and have taken into account orbital angular
momenta of quarks from lattice calculations.

\begin{figure}
\noindent \scalebox{0.8}{ \includegraphics{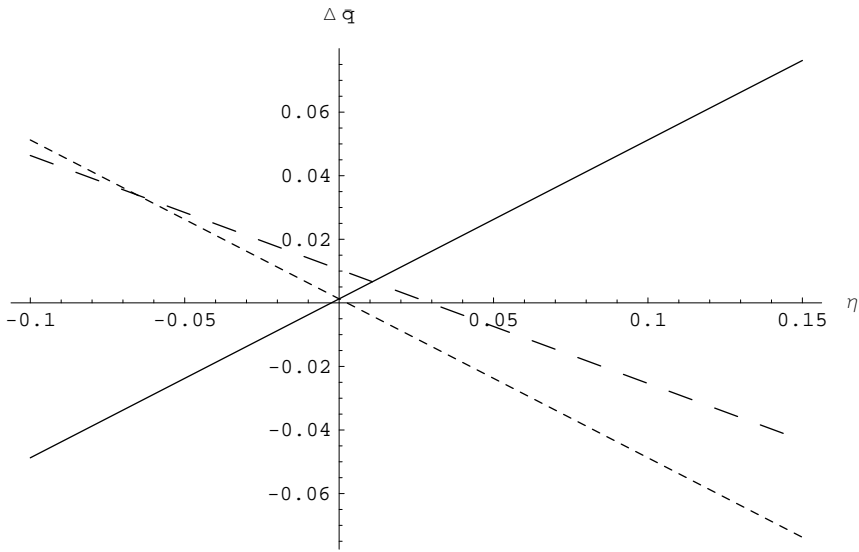}}

\caption{\label{fig4}\textit{The antiquark polarizations for
$\bar{u}$ (solid line), $\bar{d}$ (short-dashed line), and $\bar{s}$
(long-dashed line) versus $\eta$ in the model where angular momenta
of quarks are taken into account.}}

\end{figure}

 If we do not want
to use the information about $\epsilon$ from the  fit to meson
decays, we can use Eq. (19) where this parameter is eliminated and
$\Delta \bar{q}$ are functions of $\eta$. The dependence of $\Delta
\bar{q}$ on $\eta$ for u, d, and s antiquarks is shown in
Fig.~\ref{fig4}.

If we knew precisely the value of $\eta$ we could predict $\Delta
\bar{u}$, $\Delta \bar{d}$, and $\Delta \bar{s}$ values. From Eq.
(19) we see that $\Delta \bar{u}$ and  $\Delta \bar{d}$ do not
depend on orbital angular momenta of quarks; only $\Delta \bar{s}$
does. It means that precise determination of $\Delta \bar{s}$ could
be the additional test of the importance of orbital angular momenta
of quarks. When we use the $\eta$ value obtained in the HERMES
experiment we will get $\Delta \bar{u}$ and  $\Delta \bar{d}$ as in
Eq. (20) and $\Delta \bar{s} = - 0.007 \pm 0.04$. We cannot expect
additional experimental information about magnetic moments of
quarks. In the future more precise measurements of antiquark
polarizations could be a real verification of our model.

We also want to give a comparison of two models, i.e., without and
with orbital angular momenta of quarks taken into account. The
relation that follows from the COMPASS measurement of $\Gamma_{V}$,
Eq. (15), could be rewritten in the form
\begin{equation}
\eta =\frac{1}{2}a_{3}-\frac{3r}{2}
\frac{\Gamma_{V}+L_{u}+L_{d}}{f(\epsilon)+1}+\frac{1}{2}(L_{u}-L_{d}).
\end{equation}

\begin{figure}
\noindent \scalebox{0.8}{ \includegraphics{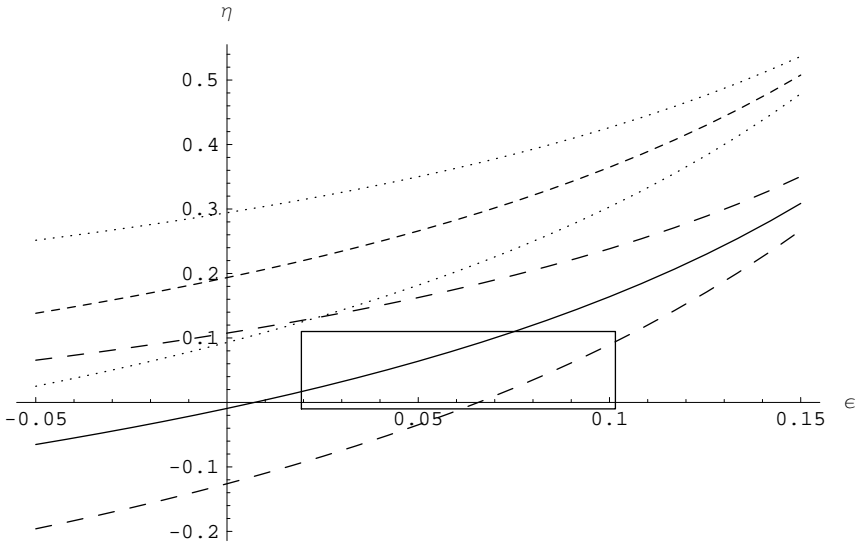}}

\caption{\label{fig5}\textit{The parameter $\eta$ versus $\epsilon$
in the model where angular momenta of quarks are taken into account
(solid line) and corresponding errors (long-dashed lines) compared
to curves in the model where angular momenta are neglected
(short-dashed line) and corresponding errors (dotted lines). The
size of the rectangle is determined by the error of $\epsilon$ from
radiative vector meson decays and the error in the measurement of
$\eta$ in the HERMES experiment.}}

\end{figure}

In Fig.~\ref{fig5} we show the function $\eta(\epsilon)$ for both
models with the corresponding errors. The rectangle is given by the
errors (1 standard deviation) of $\epsilon$ and $\eta$. It seems
that it is an indication in favor of including nonzero orbital
angular momenta of quarks.

\section{Conclusions}
From the COMPASS measurement of $\Gamma_{V}$ we have a relation
between two parameters not determined in our previous fit to
magnetic moments of baryons. Hence, we have only one independent
parameter and it could be either $\epsilon$ or $\eta$. We have
discussed two possibilities: the first with inclusion of the orbital
angular momenta of quarks suggested by calculations on the lattice
and the second with such angular momenta neglected. We have
presented in both cases the dependence of antiquark polarizations on
a single independent parameter (being $\epsilon$ or $\eta$). The
results are plotted in Fig.~\ref{fig1} to ~\ref{fig4}. In order to
find the antiquark polarizations we can use the result for $\eta$
from the HERMES experiment or the value of $\epsilon$ obtained from
the fit to radiative vector meson decays. Unfortunately the errors
are big and the results are not very conclusive.

The relation between $\eta$ and $\epsilon$  plotted with errors in
Fig.~\ref{fig5} shows that the solution with orbital angular momenta
of valence quarks taken into account is preferred.

With orbital angular momenta taken into account and the value of
$\epsilon$ taken from the fit to radiative vector meson decays, we
obtain values of antiquark polarizations. Such a procedure gives the
prediction $\eta =0.08 \pm 0.09$ that seems to be consistent with
the value $0.05 \pm 0.06$ from the HERMES experiment. Because it is
difficult to get additional information on magnetic moments of
quarks, it seems that more precise values of antiquark polarizations
(maybe from a Jefferson Lab experiment) could be a real verification
of our model.


\newpage

\begin{thebibliography}{99}
\bibitem{jb} J.Bartelski, S.Tatur,  Phys. Rev. D {\bf 71}: 014019 (2005);
\bibitem{jbr} J.Bartelski, R.Rodenberg,  Phys. Rev. D {\bf 41}, 2800 (1990);
\bibitem{jbr1} J.Bartelski, R.Rodenberg, Z. Phys. C {\bf 46}, 263 (1990);
\bibitem{bhy} A.Buchmann, E.Hern\'{a}ndes, k.Yazaki, Nucl. Phys. A {\bf 569}, 661 (1994);
\bibitem{thom3} S.Th\'{e}berge, A.W.Thomas,  Phys. Rev. D {\bf 25}, 284
(1982);  Nucl. Phys. A {\bf 393}, 252 (1983);
\bibitem{myh} H. H\o gasen, F.Myhrer,  Phys. Rev. D {\bf 37}, 1950 (1988);
\bibitem{ju}  Z.Ye, in Proceedings of the XIVth International Workshop on Deep
Inelastic Scattering Tsukuba (Japan), April 20-24, 2006 (World
Scientific, Singapore, 2007); M. Mazous {\em et al.} [Jefferson Lab
Hall A Collaboration] Phys. Rev. Lett. {\bf 99}, 242501 (2007);
\bibitem{fr1} J.Franklin,  Phys. Rev. D {\bf 66}, 033010 (2002);
\bibitem{fr2} J.Franklin,  Phys. Rev. D {\bf 30}, 1542 (1984);
\bibitem{comp} M.Alekseev {\em et al.} [COMPASS Collaboration]  Phys. Lett. B {\bf 660}, 458 (2008);
\bibitem{pdg} C.Amsler {\em et al.} (Particle Data Group), Phys. Lett. B {\bf 667}, 1 (2008);
\bibitem{comp1} V.Y.Alexakhin {\em et al.} [COMPASS Collaboration]  Phys. Lett. B {\bf 647}, 8 (2007);
\bibitem{her} A.Airapetian {\em et al.} (HERMES Collaboration), Phys. Rev. Lett. {\bf 92}, 012005
(2004); A.Airapetian {\em et al.} (HERMES Collaboration), Phys. Rev.
D {\bf 71}: 012003 (2005);
\bibitem{vec} D.A.Geffen, W.Wilson Phys. Rev. Lett. {\bf 44}, 370
(1980);
\bibitem{gloz} L.Ya.Glozman, C.B.Lang and M.Limmer Few Body Syst. {\bf 47},
91 (2010);
\bibitem{hag} Ph. H\"{a}gler {\em et al.} [LHPC Collaboration] Phys. Rev. D {\bf 77}:094502 (2008);
\bibitem{Ji} X.Ji, J.Tang and P.Hoodbhoy,  Phys. Rev. Lett. {\bf 76}, 740 (1996);
\bibitem{thom1} A.W.Thomas,  Phys. Rev. Lett. {\bf 101}, 102003 (2008);
\bibitem{thom2} A.W.Thomas,  Int. J. Mod. Phys. E {\bf 18}, 1116 (2009);
\bibitem{wak} M.Wakamatsu arXiv: 0908.0972 [hep-ph]
\end{thebibliography}
\end{document}